\documentclass[12pt,dvips]{article}
\usepackage{amscd,verbatim}
\usepackage[all]{xy}
\usepackage{graphicx}

\usepackage{amsmath}
\usepackage[psamsfonts]{amssymb}
\usepackage{amsthm}
\usepackage{euscript}

\usepackage{latexsym}

\renewcommand{\(}{\begin{equation}}
\renewcommand{\)}{end{equation} \vspace{-.05in}\linebreak}

\newcounter{saveeqn}
\newcounter{savealpheqn}

\newcommand{\alpheqn}{\setcounter{saveeqn}{\value{equation}}%
  \stepcounter{saveeqn}\setcounter{equation}{0}%
  \renewcommand{\theequation}{\mbox{\arabic{section}.\arabic{saveeqn}
\alph{equation}}}
  \renewcommand{\)}{\end{equation}}}
\def\part#1{\frac{\partial}{\partial{#1}}}%
\def\group#1{\refstepcounter{equation}\setcounter{saveeqn}{\value{equati
on}}%
  \label{#1}\setcounter{equation}{0}%
\renewcommand{\theequation}{\mbox{\arabic{section}.\arabic{saveeqn}
\alph{equation}}}
  \renewcommand{\)}{\end{equation}}}
\newcommand{\reseteqn}{\setcounter{equation}{\value{saveeqn}}%
  \renewcommand{\theequation}{\arabic{section}.\arabic{equation}}%
  \renewcommand{\)}{\end{equation}}}

\newcommand{\aalpheqn}{\setcounter{saveeqn}{\value{equation}}%
  \stepcounter{saveeqn}\setcounter{equation}{0}%
  \renewcommand{\theequation}{\mbox{
        \Alph{subsection}.\arabic{saveeqn}\alph{equation}}}
   \renewcommand{\)}{\end{equation}}}
\newcommand{\areseteqn}{\setcounter{equation}{\value{saveeqn}}%
  \renewcommand{\theequation}{\Alph{subsection}.\arabic{equation}}%
  \renewcommand{\)}{\end{equation}}}

\renewcommand{\thefootnote}{\alph{footnote}}
\renewcommand{\(}{\begin{equation}}
\renewcommand{\)}{\end{equation}}
\newcommand{\ba}{\begin{eqnarray}}
\newcommand{\ea}{\end{eqnarray}}

\newcommand{\bp}{\mathop{\vtop{\ialign{##\crcr
   $\hfil\displaystyle{}\hfil$\crcr\noalign{\kern-13pt\nointerlineskip}
   \BIG{(}\hskip0pt\crcr\noalign{\kern3pt}}}}}
\newcommand{\cbp}{\mathop{\vtop{\ialign{##\crcr
   $\hfil\displaystyle{}\hfil$\crcr\noalign{\kern-13pt\nointerlineskip}
   \BIG{)}\hskip0pt\crcr\noalign{\kern3pt}}}}}
\newcommand{\pa}{\mathop{\vtop{\ialign{##\crcr

$\hfil\displaystyle{\oplus}\hfil$\crcr\noalign{\kern+1pt\nointerlineskip
}
   \hspace{.08in}$^{\alpha=0}$\hskip6pt\crcr\noalign{\kern3pt}}}}}

\newcommand{\R}{\ensuremath{\mathbb R}}

\newcommand{\FF}{\ensuremath{\mathbb F}}

\newcommand{\cE}{\ensuremath{\mathcal E}}

\newcommand{\cN}{\ensuremath{\mathcal N}}

\newcommand{\C}{\ensuremath{\mathbb C}}

\newcommand{\Z}{\ensuremath{\mathbb Z}}

\newcommand{\beq}{\begin{equation}}
\newcommand{\eeq}{\end{equation}}

\numberwithin{equation}{section}

\def\hsp#1{\hspace{#1in}}

\catcode`\@=11
\def\vereq#1#2{\lower3pt\vbox{\baselineskip1.5pt \lineskip1.5pt
\ialign{$\m@th#1\hfill##\hfil$\crcr#2\crcr\sim\crcr}}}
\catcode`\@=12

\makeatletter
\newcommand\figcaption{\def\@captype{figure}\caption}
\newcommand\tabcaption{\def\@captype{table}\caption}
\makeatother
\renewcommand{\(}{\begin{equation}}
\renewcommand{\)}{\end{equation}}

\newcommand{\CC}{{\mathbb C}}
\newcommand{\RR}{{\mathbb R}}
\newcommand{\ZZ}{{\mathbb Z}}
\newcommand{\QQ}{{\mathbb Q}}

\theoremstyle{plain}

\theoremstyle{definition}

\newcommand{\CP}{\CC \text{P}}

\begin{document}

\begin{titlepage}
\begin{flushright}
hep-th/0511087\\
NSF-KITP-05-94
\end{flushright}

\vspace{2em}
\def\thefootnote{\fnsymbol{footnote}}

\begin{center}
{\Large\bf  The Elliptic curves in gauge theory, string theory,\\
\vspace{2mm}
and cohomology}
\end{center}  
\vspace{1em}
\begin{center}
{\large Hisham Sati} \footnote{E-mail: \tt
hisham.sati@adelaide.edu.au}
\end{center}

\begin{center}
\vspace{1em}  
{\em  
Department of Theoretical Physics\\
Research School of Physical Sciences and Engineering\\
The Australian National University\\
Canberra, ACT 0200\\
Australia\\

\hsp{.3}\\  
  Department of Pure Mathematics\\   
       University of Adelaide\\
       Adelaide, SA 5005,\\
       Australia}\\

\end{center}
       
\vspace{0em}

\begin{abstract}
\noindent 
Elliptic curves play a natural and important role in
elliptic cohomology. In earlier work with I. Kriz,  
these elliptic curves were interpreted physically in two ways: as corresponding to the 
intersection of M2 and M5 in the context of (the reduction of M-theory to) type IIA and as 
the elliptic fiber leading to F-theory for type IIB. 
In this paper we elaborate on the physical setting for various generalized 
cohomology theories, including elliptic cohomology, and we note that the above 
two seemingly unrelated descriptions can be unified using Sen's 
picture of the orientifold limit of F-theory compactification on $K3$, which unifies the 
Seiberg-Witten curve with the F-theory curve, and through which we naturally explain the constancy of the modulus that emerges from elliptic cohomology.  
This also clarifies the orbifolding performed in the previous work and 
justifies the appearance of the $w_4$ condition in the elliptic refinement
of the mod 2 part of the partition function. We comment on the cohomology theory 
needed for the case when the modular parameter varies in the base of the elliptic 
fibration.
\end{abstract}

\vfill

\end{titlepage}
\setcounter{footnote}{0}
\renewcommand{\thefootnote}{\arabic{footnote}}

\pagebreak
\renewcommand{\thepage}{\arabic{page}}

\section{Introduction}
The form-fields of string theory and M-theory are important in the study of the global structure of the these theories.  
In previous work with I. Kriz \cite{KS1,KS2,KS3}, elliptic cohomology theory 
was proposed to describe the fields of type II string theories and their partition functions, as refinements of the K-theoretic description of the fields \cite{MW,FH} and of the partition function \cite{DMW}.
In the twisted case \cite{MS}, i.e. 
in the presence of the NSNS field $H_3$, the twisted K-theoretic description is discussed in
\cite{MS} for type IIA, and an S-duality covariant description for type IIB using generalized cohomology 
refinements was proposed  in \cite{KS2}.
For M-theory, in \cite{S1,S2,S3}, a higher degree analog of K-theory was proposed. In both string theory and M-theory, several generalized cohomology theories were considered, including the theory of topological modular forms,
TMF \cite{AHS}.

\vspace{3mm}
The type IIA K-theoretic partition function was constructed and matched by Diaconescu, Moore and Witten \cite{DMW} with the Rarita-Schwinger and form-field part of the partition of M-theory \cite{W1} . The cancellation of the DMW anomaly given by the seventh Stiefel-Whitney class  
$W_7$ led to the interpretation \cite{KS1} as a condition of orientation in elliptic cohomology. Consequently, 
an elliptically-refined partition function was proposed. In \cite{KX} part of this conjecture was verified. Elliptic cohomology comes equipped with an elliptic curve, defined over the coefficient ring of the theory, usually an integral polynomial in certain generators, identified as higher degree analogs of the Bott generator.  Since elliptic cohomology appeared from rather physical considerations, namely the cancellation of an anomaly, it is expected that 
the structures that come with elliptic cohomology should correspond to some physical interpretation.
For theories with one step higher
\footnote{in the so-called chromatic tower which will be explained in the next section.}
 than K-theory, the generators have dimension two and six, a fact which 
was interpreted in \cite{KS1} as corresponding, respectively,  to the string and the NS5-brane, or to the 
M2-brane and the M5-brane in the M-theory limit. The elliptic curve of
elliptic cohomology  was consequently interpreted as corresponding to the elliptic curve arising from the intersection of the M2-brane and the M5-brane. This system reduces to strings on D-branes and so 
is natural to consider from the string theory point of view (see \cite{Str,Town}).
The anomalies of such a system have been studied in \cite{DFM}. 

\vspace{3mm}
On the other hand, an S-duality covariant approach to describe the fields of type IIB string theory was proposed in \cite{KS2,KS3}. In studying modularity in \cite{KS3}, the elliptic curve of elliptic cohomology was given another interpretation, namely as corresponding to the fiber of F-theory over type IIB string theory. TMF leads to anomaly-free modularity. However, this theory is not an elliptic cohomology theory, but is rather obtained from it by orbifolding, in the same way that one would get real K-theory $KO$ from complex K-theory $K$. As the elliptic curve was taken to correspond to a `physical' space, i.e. the elliptic fiber in F-theory, the orbifolding was also proposed to correpond to 
actual physical orbifolding in spacetime, as the one leading from type IIB string theory to type I theory.

\vspace{3mm}
However, several points remain to be clarified. First of all, the above two descriptions seem to be a priori unrelated, and even possibly incompatible. For instance, why would one expect that the elliptic fiber of F-theory
would have anything at all to do with the intersection of M-branes in M-theory. Second, the description in terms of F-theory requires the modular parameter $\tau$ to be a field that varies in some base space of an elliptic fibration, whereas the modular parameter that appears in elliptic cohomology, and in TMF, is simply a constant complex number. Third, the construction of the mod 2 part of the elliptic partition function in \cite{KS1} required the spacetime to be oriented with respect to $EO$, i.e. that the Stiefel-Whitney class $w_4$ vanishes, a condition which is more natural in type I and heterotic string theories.
Fourth, the `physical' elliptic curves are of course defined over $\CC$, while the cohomology curve is defined over the coefficient ring of the theory. The aim of this note is to reconcile the above points among themselves, to make the previous picture in \cite{KS3} compatible with the web of dualities, and to propose further generalizations.

\vspace{3mm}
For the first three items, we use Sen's description \cite{Sen1,Sen2,Sen3} of the compactification of F-theory on $K3$ in the orbifold limit in the moduli space where the surface appears as an orbifold of a four-torus. This automatically justifies the orbifolding done in \cite{KS3} and makes the picture compatible with the web of dualities. The intersection of M2 and M5 in M-theory is a self-dual string, and such a configuration corresponds to the lift \cite{sol} of Seiberg-Witten theory \cite{SW1,SW2}, i.e. $\cN =2$ $d=4$ super-Yang-Mills theory, to M-theory. Sen also provided the embedding of Seiberg-Witten theory in F-theory using the same orbifold construction, where the mathematical resemblance of the two theories is striking and it is interesting that such a connection involves properties of the elliptic curves. The gauge theory is $\cN=2$ SYM with group $SU(2)$ and with four quark flavors. The moduli space of this theory was characterized by a gauge-invariant quantity $u$, and the complex coupling constant $\tau$ varies as we move in the $u$-plane \cite{SW1,SW2}. Sen \cite{Sen1} has shown that this is identical to the
 F-theory background with $u$ labelling the coordinates of the base
of the elliptic fibration and $\tau$ denoting the axion-dilaton modulus.
The particular orientifold background above corresponds to the classical
limit, which is singular since $Im(\tau)$ becomes singular in some regions.  
The identification is as follows
\begin{eqnarray}
\nonumber
{\rm gauge~ coupling} &\leftrightarrow& {\rm axion-dilaton}
\nonumber\\
{\rm masses}~ m_i ~{\rm of~ the~ hypermultiplets} &\leftrightarrow& 
{\rm positions~ of~ 
7-branes}
\nonumber\\
{\rm the~ vev}~ z=\langle {\rm tr}~ \phi^2 \rangle {\rm ~of~ the~ adjoint~ scalar} 
&\leftrightarrow& {\rm a ~point~ in~ the~ base~ (locally}~ \RR^2/\ZZ_2).
\end{eqnarray}

\vspace{3mm}
We start with a section on the general setting for generalized cohomology theories encountered in this paper and in the previous work. In particular we discuss the refinement of the Ramond-Ramond fields, (Hodge or electric/magnetic) duality symmetry and how it fits in this formalism, and the dynamics of M2- and M5-branes and their intersections. Part of this interpretation has appeared in \cite{S3}. We then review the description of the three elliptic curves and then relate them by Sen's orientifold pictures, making the required clarifications along the way and 
in the discussion section. Sen's picture was made more than a mathematical correspondence in \cite{BDS}, where it was shown that the picture can be captured by a probe D3-brane, representing the gauge theory, in the presence of D7-branes. We comment on this and on the consequence of deformation away from the limit
to variable $\tau$ for the corresponding cohomology theory. We also interpret the formal parameter $q$ as a coupling expansion parameter. While the arguments in this paper are mostly  qualitative, we do provide in the discussion section some proposals for quantitative descriptions and further checks.

\section{The physical setting for generalized cohomology}

In this section we argue for a physical setting for various generalized 
cohomology theories by studying the structures involved.
We look at the fields that appear in string theory and M-theory and identify
which aspects of the fields are described by which cohomology theory.
There are two issues: RR fields getting information about an 
elliptic curve, i.e. an elliptic refinement of the K-theoretic description, and, second, 
Hodge duality between the string and the NS5-brane or between the M2-brane and the 
M5-brane. The second case involves complex-oriented generalized cohomology theories,
i.e. ones descending from complex cobordism. It is important to distinguish elliptic 
vs. non-elliptic theories.

\subsection{Interpreting the ``generalized" in generalized cohomology} 
A {\it generalized} cohomology theory is a theory that satisfies all 
the properties of a usual cohomology theory except for one, which is 
called the dimension axiom. This is the cohomology of a point, i.e.
${\mathcal H}^*(pt)$. This is also called the coefficient ring of the 
theory. For `un-generalized' theories, this is just $\ZZ$. For the 
generalized theories this can be something much more complicated, e.g.
a formal polynomials in certain parameters, with coefficients that can 
be integral or integral mod $p$.

\vspace{3mm}
Among the generalized cohomology theories in which we are interested and
which have appeared in the study of type II string theories \cite{KS1,KS2,KS3} -- and 
to some extent also in M-theory \cite{S1,S2,S3} --
are the theories in the so-called chromatic 
tower of spectra,
\footnote{ For
our purposes, the word ``spectrum'' simply means a
generalized cohomology theory.}
i.e. the ones that descend from complex cobordism 
theories. Those can be seen, looking from the other end, as a 
generalization of K-theory by alowing Bott generators of higher dimensions.
An example of this type is (integral) Morava K-theories. 
In the previous work and in this paper, we also encounter a generalization 
of another type, namely {\it elliptic} cohomology theories which have an 
{\it elliptic} structure manifested by an elliptic curve parametrizing the
coefficients in the cohomology of a point.  
This kind of cohomology theories will still contain the usual 
Bott element $v_1=u$ of dimension two, but will have coefficients that
are parametrized by the elliptic curve.

\vspace{3mm}
What is the implication of this for the physics? We will give some 
intuitive arguments on what the above means physically for us. 
We would like to interpret the fact that we have a nontrivial coefficient 
ring, which is the cohomology of a point, as indicating a nontrivial 
structure on (or over) that point. For example, this could be a
generalization of a bundle structure. From the algebraic side, this
is just having a sheaf structure over a given point. For us, this is
close to what a scheme is, and so we see 
that this will perhaps 
justify the
use of the concept of a {\it generalized elliptic curve} in such cohomology
theories.
The above implies that a point in this context has more structure than 
what one would usually associate to a point. In the particular case of 
elliptic cohomology, this structure will be a (family of generalized) 
elliptic curves. So, again intuitively, we have a somewhat hidden  
elliptic curve over each point. This is very similar to the Kaluza-Klein 
idea, in the sense that we see an internal structure over each point. 
This is yet another way of justifying the interpretation of the elliptic 
curve
appearing in elliptic cohomology as corresponding to a physical elliptic 
curve in \cite{KS1,KS3}.

\subsection{Physical association of cohomology theories} 

We start by looking at a special class of generalized cohomology theories,
namely elliptic spectra. An {\it elliptic spectrum} consists of \cite{AHS}:
\begin{enumerate}
\item an even, periodic, homotopy commutative ring spectrum $E$ with
formal group $P_E$ over the coefficient $E^0(pt)$;
\item a generalized elliptic curve $\mathcal{E}$ defined over $E^0(pt)$;
\item an isomorphism $t$ of $P_E$ with the
formal completion $\hat{\mathcal{E}}$ of $\mathcal{E}$.
\footnote{Note that an elliptic curve is a one-dimensional abelian group.}

\end{enumerate}
The periodicity condition is explained by the following definition. A {\em 
$2$-periodic}
ring spectrum as a generalized cohomology theory $E$ with an orientation 
of
the complex line bundle on $\C P^{\infty}$ which, when
restricted
to $\C P^1$, has an inverse in $E_*$. The {\it evenness} condition 
means that all odd cohomology vanishes, i.e. $E_{2n+1}=0$ for all
$n$.
This definition involves the notion of {\it generalized elliptic curves} 
over a ring $R$. This means a marked curve over the scheme ${\rm Spec}(R)$
\footnote{Intuitively, this is just a `thickened' version of the original curve
where the thickening can be understood as the structure over the points, 
(e.g. multiplicities). Supermanifolds, in one version, have such a 
description.}
which is 
locally isomorphic to a Weierstrass
curve, i.e. 
\(
\mathcal{E} :~y^2+a_1xy+a_3y=x^3+a_2x^2+a_4x+a_6,
\label{Weie}
\)
with $a_1,a_2,a_3,a_4,a_6 \in R$.
We will come back to the discussion of the elliptic curve and formal groups 
in the next section.

\vspace{3mm}
{\bf Ramond-Ramond fields}:

\vspace{2mm}
First we argue that, granted a justification for its use, 
the concept of an elliptic spectrum is not such a 
strange 
concept for physics. It is in
fact already closely related to the discussion of Ramond-Ramond ($={\rm RR}$) fields in
K-theory. Recall that
in the K-theoretic description of the RR fields 
\cite{MW,FH}--
say in type IIA-- one has the total gauge-invariant field strength
written as ${F} = \sum_{n=0}^{5} {{F}}_{2n}$. In order to make the
RR field strengths homogeneous of degree zero, one can \cite{F} use
K-theory with coefficients in $K(pt)\otimes \R \cong \R [ [v_1,
v_1^{-1}]]$, where the inverse Bott element $v_1\in K^2(pt)$   
has degree $2$; then the total RR field strength
is written as the total uniform degree-zero expression
\(
F=\sum_{m=0}^5 (v_1)^{-m}F_{2m}.
\label{rr}\)
Strictly speaking, we are dealing with cohomology rather than K-theory
because we have already taken the image with respect to the Chern
character included within the $F$'s (see \cite{MS} or \cite{S3}).

\vspace{3mm}
Let us now rewrite this in a way that {\it resembles} an elliptic spectrum 
\cite{AHS}.
First note that there are three kinds of algebraic groups of dimension 
one: the additive
group ${\mathbb{G}_a}$, the multiplicative group ${\mathbb{G}_m}$, and
elliptic curves. The first one is just the complex plane $\CC$ with the
operation of addition on complex numbers and the second is $\CC^*$ with
the multiplication operation. An elliptic curve over $\CC$ is of the form 
$\CC/\Lambda$ for some lattice
$\Lambda \subset \CC$. The map of formal groups derived from $\CC
\rightarrow \CC/\Lambda$ gives an isomorphism $t_{\Lambda}$, from the  
additive formal group $\hat{\mathbb{G}}_a$ to the formal completion
of the elliptic curve ${\hat{\mathcal{E}}}_{\Lambda}$. Let $R_{\lambda}$ be the
graded ring $\CC[u_{\Lambda},u_{\Lambda}^{-1}]$ with $|u_{\Lambda}|=2$,
and define an elliptic spectrum $H_{\Lambda}=(E_{\Lambda}, {\mathcal{E}}_{\Lambda},
t_{\Lambda})$ by taking $E_{\Lambda}$ to be the spectrum representing
$H_*(-;R_{\Lambda})$, ${\mathcal{E}}_{\Lambda}$ the elliptic curve $\CC/{\Lambda}$,
and $t_{\Lambda}$ the isomorphism described above. So if we identify
$v_1$ with $u_{\Lambda}$ then in describing the RR fields as above
we are dealing with an analog of the elliptic spectrum. Strictly speaking,
it is not elliptic since this would require $n=2$.

\vspace{3mm}
Among the examples given in \cite{AHS} and used in \cite{KS1} is the 
elliptic spectrum
\(
E_*=\Z[a_1,a_2,a_3,a_4,a_6][u,u^{-1}]
\label{E}
\)
associated with the Weierstrass curve which we write in the form
\(
y^2+a_1 uxy +a_3 u^3 y = x^3 +a_2 u^2 x^2 +a_4 u^4 x +a_6
u^6,\)
where ${\rm dim}(x)=4$ and ${\rm dim}(y)=6$.
Again, this connects with the RR fields since the coefficient ring 
(\ref{E}) has $u$ but yet also has
information about the elliptic curve
\footnote{which was interpreted in \cite{KS3} as corresponding to the F-theory elliptic fiber,
the further justification of which is one of the aims of this note.}
, since over each point we have
such a fiber. Thus this is the most straightforward generalization of 
the above K-theoretic homogeneous degree zero description for the
RR fields in type II string theory. We take this to correspond to 
refinements \cite{KS3} leading to the RR formula
\(
F(x)=\sigma(X)^{1/2}ch_E(x),
\label{RR}
\)
where $\sigma(X)$ is the Witten genus of the manifold $X$ and $ch_E$ is the
Chern character of the theory $E$. This still has the same Bott generators as 
K-theory, but now instead of taking coefficients that are integers, we are taking 
them to be integral polynomials in the coefficients $a_i$ of the Weierstrass curve.

\vspace{3mm}
{\bf Electric/Magnetic duality:}

\vspace{2mm} 
Among the interesting complex-oriented generalized cohomology theories 
obtained from complex cobordism (see \cite{KS1} for an exposition)
are Johnson-Wilson theory $BP\langle n \rangle$, Landweber elliptic cohomology 
 $E(n)$, Morava 
K-theory $K(n)$ and integral Morava K-theory ${\widetilde K}(n)$. For $n=2$, the 
coefficient rings of these and of complex K-theory are given as
\begin{eqnarray}
BP_*&=&\ZZ[v_1,v_2],
\nonumber\\
E(2)_*&=&\ZZ[v_1,v_2,v_2^{-1}],
\nonumber\\
{\widetilde K}(2)_*&=&\ZZ[v_2,v_2^{-1}],
\nonumber\\
{K}(2)_*&=&\ZZ/p[v_2,v_2^{-1}],
\nonumber\\
K_*&=&\ZZ[v_1,v_1^{-1}].
\label{coeff}
\end{eqnarray}
Such a `hierarchy' of theories is sometimes referred to as (part of) the 
chromatic tower for spectra. The elements in the coefficients are related
to the formal group laws. For
$K(n)_*$ this has {\em height $n$}, which means that, in particular for
$p=2$ and $3$ respectively,
\begin{eqnarray}
{[2]}_F x &=& x +_F x=v_n x^{2^n}
\nonumber\\
{[3]}_F x &=& x +_F x +_F x=v_n x^{3^n}.
\label{fg}
\end{eqnarray}

\vspace{3mm}
We would like to comment on the structure of the
above theories and propose their implication for physics.
For $p=2$, the dimensions of the generators are $2$ and $6$ for $v_1$ and 
$v_2$ respectively. In \cite{KS1}, ${\widetilde K}(2)$ and 
$E(2)$ were used, and there it was proposed that the generators 
correspond to worldvolumes of the corresponding intersection of membranes
and fivebranes. From a IIA point of view, it was proposed
that this was related to the fundamental string and the
Neveu-Schwarz fivebrane, respectively. 
\footnote{Strictly speaking, the following discussion for $p=2$ is better adapted 
for string theory, and in M-theory one might need to go \cite{S3} to $p=3$, or 
better yet, to TMF where all 
primes are treated democratically. In any case, the discussion 
can be made to be generic.} 

\vspace{3mm}
Here we would like to go further and propose the setting for the above
theories in string theory and M-theory. Since $v_1$ corresponds to the
(boundary of the) membrane and $v_2$ corresponds to the fivebrane, and since 
those in turn couple to the field $C_3$ and to the potential $C_6$ of the dual of 
its field strength, respectively, then the $v_i$ ($i=1,2$) should tell us
something about Hodge duality. So in some sense, generalized cohomology theories 
that contain only $v_1$ are `electric', theories that contain only $v_2$ are 
`magnetic', and theories that contain both $v_1$ and $v_2$ are 
`electro-magnetic'.
Therefore, by inspecting (\ref{coeff}), we propose the following identifications
between generalized cohomology theories and physical theories,
\begin{eqnarray}
{\rm K-theory} &\longleftrightarrow& {\rm non-dual~ theory}
\nonumber\\
{\rm (integral)~ Morava~ K-theory}  &\longleftrightarrow& {\rm dual~ theory}
\nonumber\\
{\rm Brown-Peterson~ theory}  &\longleftrightarrow& {\rm duality-symmetric~ theory}
\nonumber\\
{\rm Landweber~ theory}  &\longleftrightarrow& {\rm special~ duality-symmetric~ theory.}
\nonumber
\end{eqnarray}

\noindent The last one is termed ``special'' because of the fact that the second generator, $v_2$, 
is inverted while 
the first one, $v_1$, is not, and so this suggests a more privileged  role for the
dual field. In light of the identifications we have done, namely the brane worldvolumes with
the generators, and the generalized cohomology theories with types of physical theories 
with respect to duality, we further propose that K-theory is a theory that has to do with membranes
(or strings, or electric fields), Morava K-theory with fivebranes (or magnetic fields),
Brown-Peterson theory with membranes and fivebranes, and Landweber's theory with 
membranes and fivebranes and where the latter play a special role since the generator 
corresponding to them, $v_2$, is inverted.  
One way to explain this inversion is to say that the M5-brane and the NS5-brane 
are more fundamental that the M2-brane and the string. This might be justified by either 
a form of the Hanany-Witten effect \cite{Han} adapted for this case, where, for example, 
an M2-brane 
is created out of an M5-brane  
\footnote{Some discussion on this can be found in \cite{ES, S3}.}
or by saying that fivebranes are instinsically more fundamental than either the membrane 
or the string.

\vspace{3mm}
There are two aspects to explain. First, that D-branes and strings can be obtained from
the M-branes, and second, that the M5-brane in some sense already encodes the M2-brane.
Since M-theory does not have a perturbative coupling constant analogous to the dilaton
in string theory, the relationship between M2 and M5 (and at the level of fields, 
between $C_3$ and $C_6$) should be different from the usual strong-weak coupling 
duality between strings and NS5-branes in ten dimensions. Based on this, it was argued in
\cite{DVV} that the M2-brane may be a limiting case of the M5-brane, which would
imply that M-theory is self-dual. One way to see this is to note that \cite{DVV} 
besides the dual
six-form $C_6$ of eleven-dimensional supergravity, the
five-brane also couples directly to the three-form $C_3$ itself, via
an interaction of the type
\(
\label{CT}
\int C_3\wedge T_3,
\)
where $T_3$ is a self-dual three-form field strength that lives on the
M5-brane world-volume. 
Maps between the generalized cohomology theories are characterized by the
corresponding maps, if they exist, between the coefficient rings.
There are maps from $BP$ to all other descendents, and also
from $BP\langle n\rangle$ to $E(n)$, but there are no maps between the 
$K(n)$'s for different $n$.
This has implications on duality, because for us this implies that
one cannot relate $K(1)$ information, i.e. usual K-theoretic information
from $v_1$, to the `dual' $K(2)$ information, i.e. the Morava K-theoretic
information from $v_2$, of course unless we are already in the context 
of a bigger theory that contains bothpowers

\vspace{3mm}
{\bf M2-M5 states and interactions:}

\vspace{2mm}
By allowing this field $T_3$ to have non-trivial fluxes 
through the three-cycles on the world-brane, the five-brane can thus  
in principle carry all membrane quantum numbers. These configurations 
are therefore naturally interpreted as bound states between the two
types of branes. Based on this the authors of \cite{DVV} argue 
that the M5-brane is a natural
candidate to give a more unified treatment of all BPS states in string
theory. Indeed, they  give a unified description of all BPS states of M-theory compactified
on $T^5$ in terms of the M5-brane.
An important ingredient in this
formalism is the idea that the relevant degrees of freedom on the   
five-brane are formed by the ground states of a string theory living
on the world-volume itself.
In \cite{DVV} it was further argued that by comparing to the analysis of D-brane states
\cite{RR}, 
it should be possible
to make concrete
identification between specific five-brane excitations and
configurations of D-branes and fundamental strings. 
Via this correspondence our arguments on the BPS spectrum
should also give useful information about the bound states of strings and
D-branes \cite{Bound}.

\vspace{3mm}
What we are advocating is that the generalized 
cohomology theories that we are dealing with have the correct mathematical 
structure to incorporate the interactions between M2-branes and M5-branes,
which, in light of the arguments in \cite{DVV}, means that this also incorporates
bound states and interaction of type IIA string theory (and type IIB 
by T-duality).
In particular, this suggests identifying the parameter $q$ that appeared in 
the refinement of the type IIA and M-theory partition functions in \cite{KS1}
as a coupling constant which encodes the interactions of M2-M5 systems. Then, this 
would also imply that this coupling constant also encodes the interactions
of string and branes in ten dimensions.

\section{ The elliptic curve in Elliptic cohomology}
\label{ell}

We start with discussing the (in-)dependence of the generalized cohomology
theories on the primes $p$. In string theory, usually the prime $2$ is important, 
i.e 2-torsion appears in the fields, and this can be seen for example in the $K$-theory 
calculation of the IIA partition function in \cite{DMW}. A notation like 
$\ZZ[\frac{1}{2}]$ means that 2 is inverted, which implies
that we stay away from 2-torsion. In contrast, when we say we localize at  
2, this means that we are looking at 2-torsion, i.e. we cannot freely     
divide by 2. Then, from the point of view of the chromatic tower, this means
that we are looking at $K(1)$-local information since this contains $v_1$.
Of course the same analysis holds for the other primes. There is a simplification
that occurs in the Weierstrass equation (and the definition of $E$), but that is
at the expense of losing torsion information since it requires inverting
of $2$ and $3$. 

\vspace{3mm}
For $n=2$ in (\ref{fg}), one can also consider cohomology theories whose formal group
laws are elliptic, i.e. are obtained by Taylor expansion of the group law on an
elliptic curve over some commutative ring. Such theories are the {\em
complex-oriented elliptic cohomology theories}.
For the sake of exposition, let us briefly see how this works (this is standard mathematical literature, see
e.g. \cite{Hus} or \cite{Sil}).
For (\ref{Weie}), introduce new variables $t=\frac{-x}{y}$ and
$s=\frac{-1}{y}$. By iteratively solving for $s$ in terms of $t$, one can
write $x$ and $y$ in terms of the variable $t$ only, i.e.
\(
x=t^{-2}-a_1t^{-1}-a_2-a_3t-(a_4+a_1a_5)t^2 + \cdots,
\)
and $y$ is just $-x/t$. We see that $x$ and $y$ are power series
expansions in the variable $t$ and with coefficients given by the $a$'s,
and so they are in $\ZZ[a_1,a_2,a_3,a_4,a_6][[t]]$.
\footnote{A notation like $f((t))$ means power series in $t$, while $f[[t]]$ means
restriction to non-negative powers.}

\vspace{3mm}
The group law can be obtained by working near $t=0$ ($t$ identified as a
local parameter near the
origin $(t,s)=(0,0)$ in $E$). If $(t_1,s_1)+(t_2,s_2)=(t_3,s_3)$ on the
elliptic curve $E$ in the $(t,s)$-plane, then $t_3\equiv \Phi(t_1,t_2)$
has the form
\(
t_3=t_1+t_2 -a_1t_1t_2 -a_2(t_1^2t_2+t_1t_2^2) -
2a_3(t_1^3t_2+t_1t_2^3)+\cdots,
\)
and so $\Phi(t_1,t_2)$ similarly is in
$\ZZ[a_1,a_2,a_3,a_4,a_6][[t_1,t_2]]$.
If the coefficients $a_j$ of $E$ lie in a ring $R$, then
$t_3=\Phi(t_1,t_2)$ is in $R[[t_1,t_2]]$.
The formal series $\Phi(t_1,t_2)$ arising from the group law on $E$ is a
{\it formal group law} FGL. This is given by 
\(
{\hat C}(x,y)=x+y -a_1xy -a_2(x^2y+xy^2)
-(2a_3x^3y -(a_1a_2 -3a_3)x^2y^2 + 2a_3xy^3) + \cdots.
\)
The FGL can also be defined as the way line bundles behave under tensor product, 
\(
{\hat G}_E(x,y)=e(L_1\otimes L_2) \in E^*({\CP}^{\infty} \times {\CP}^{\infty})
\cong \pi_*E[[x,y]],
\)
where $e$ is the Euler class.  

\vspace{3mm}
There are various models for complex-oriented cohomology $E$
which are characterized by their coefficient rings $E^*(pt)$.
Choosing a complex-oriented elliptic cohomology theory corresponds to choosing
coordinates on the elliptic curve. The Weierstrass curve is the universal
(generalized) elliptic curve, and is
given by the equation $$y^2x+a_1 xyz +a_3 y^3=x^3+a_2 
x^2z+a_4 xz^2 +a_6 z^3$$ over the ring $\Z[a_1,a_2,a_3,a_4,a_6]$,
where the parameters $a_i$ are
(generalized) modular forms. However, this curve has automorphisms and thus
the corresponding theory with  $E_*=\Z[a_1,a_2,a_3,a_4,a_6][u,u^{-1}]$
is not universal. 

\vspace{3mm}
 {\bf Reduction of coefficients}:
\vspace{2mm}

In the physical cases, i.e. in Seiberg-Witten theory and in F-theory, the elliptic curves 
are physical parts of spacetime, i.e. are looked at as Riemann surfaces. Thus they are defined over $\CC$. On the 
other hand, the elliptic curves  that show up in elliptic cohomology are defined over various
different fields or rings. This includes finite fields $\FF_p$, integral or $p$-integral polynomial 
rings. Thus, at a first glance,  the two pictures seem to be incompatible. However, there is 
a standard procedure to get curves over the latter coefficients starting with curves over
$\CC$. \footnote{The question of whether the resulting 'discrete' varieties correspond {\it in 
reality}  to some discrete versions of spacetime seems to be a much deeper question that 
goes far beyond the scope of this paper.}

\vspace{3mm}
From an elliptic curve over $\CC$ one can get an elliptic 
curve over $\ZZ$ if one can define the curve over $\QQ$. The latter of course is not
a serious condition and is in some sense the `default' situation.
One can get  the curve over $\ZZ [x_1,x_2,\cdots]$ or $\ZZ[t]$ if one can 
define the curve over some  finite extension of $\QQ$, again not a very serious 
condition. In all the relevant cases, the resulting curve would have the same Weierstrass form, except that the coefficients $a_i$, instead of taking values in $\CC$, now take values in the new 
field or ring. For example, for $\ZZ[t]$, one has $a_i(t)\in \ZZ[t]$, and 
\(
y^2+ a_1(t) xy + a_3(t) y = x^3 + a_2(t) x^2 + a_4(t)x + a_6(t),
\)
giving an elliptic curve $E$ over $\ZZ[t]$.

\section{The elliptic curve in Seiberg-Witten theory}

For Seiberg-Witten theory \cite{SW1} \cite{SW2}, one can have any genus $g$ for the 
curve and not 
necessarily just the elliptic case $g=1$. 
The latter is the case for the originally considered  
gauge group $SU(2)$ only. In general, one can have any genus for the curve. In particular
for groups of type $A_n$, the genus is equal to $n$ \cite{KLTY, AF}. All ADE groups have also 
been considered, leading to different genera (greater than one) \cite{DS, BL, KLMVW}.
So one might ask the question of whether elliptic curves play any particularly important role in
in the $N=2$, $d=4$ gauge theory.

\vspace{3mm}
The  Seiberg-Witten gauge theory results can be obtained from type IIA string theory.
In \cite{KLMVW} 
The $N=2$ $SU(2)$ SYM can be regarded as the worldvolume
theory of parallel D4-branes. The D4's have a finite extent in one
direction, along which they end on NS5-branes.
it was shown how to obtain this genus $g$ Riemann surface $\Sigma_g$  from string theory on a 
Calabi-Yau threefold of the form $\Sigma_g \times \RR^4$, which represents a symmetric fivebrane. 
The result was that the Riemann surface is given a concrete physical meaning in \cite{KLMVW}, where
the BPS states correspond to self-dual strings (cf. \cite{DL}) that wind geodesically around the homology cycles,
viewed as boundaries of D2-branes ending on the curve part of the fivebrane. This D2-brane can be 
viewed as the the disk formed from filling the one-cycle on the Riemann surface, giving the string on 
the Riemann surface as a boundary.

\vspace{3mm}
Witten \cite{sol} provided a derivation of Seiberg-Witten from M-theory
and interpreted a geometric interpretation for the curve. The
corresponding configuration is given in terms of single M5-brane with
worldvolume $\RR^4 \times \cE$ with $\cE$ an elliptic curve embedded in the
Euclidean space $X$ spanned by $x^4,x^5,x^6,x^{10}$, where $x^{10}$ is the
eleventh direction, i.e. the would-be `M-theory circle', and $\RR^4$
spanned by $x^0,x^1,x^2,x^3$. This space is
endowed with a complex structure such that $s=x^6+ix^{10}$ and
$v=x^4+ix^5$ are holomorphic with the requirement that $\cE$ is a holomorphic curve
in $X$ of degree 2 in $v$.

\vspace{3mm}
The BPS states in the fivebrane worldvolume field theory correspond to
M2-branes ending on the M5-brane \cite{Str, Town}. The boundary of M2
has to lie on $\cE$ and it couples to the self-dual two-form on the
world-volume of M5. The mass of the BPS states is given by that
of the M2, which in turn is given by the brane tension times the area
\(
m=T_2 \int_{M2} |\omega |,
\)
where the area is given by the pullback of the holomorphic 2-form
$\omega=ds\wedge dv$ \cite{FS}.
Since $\partial M2=\cE$,
\(
m^2=T_2 \left| \int_{\partial M2} v(t)\frac{dt}{t} \right|^2
\)
where $v$ is written as a function of $t=e^{-s}$.
\footnote{written this way since the eleventh direction is a circle.}
Thus the mass of M2 reduces to an integral of a meromorphic one-form on
$\cE$, and the Seiberg-Witten differential is given by \cite{FS}
\(
\lambda_{SW}=v(t)\frac{dt}{t}.
\)
From the point of view of the self-dual string, this is just the tension
\cite{KLMVW}.

\vspace{3mm}
One gets different elliptic curves depending on whether or not there is matter, and 
on the kind of matter allowed.
After including hypermultiplets, the Coulomb branch of vacua is still
parametrized by a copy of the $u$-plane (where $u$ is
related to $\langle {\rm Tr}~ \phi^2 \rangle$ in the underlying theory),
but now the $u$-plane parametrizes a different family
of elliptic curves.  The appropriate
families, which depend on the
hypermultiplet bare masses, have the form \cite{SW1,SW2}:
\footnote{By reduction (see e.g.
\cite{Hus}) the curve (\ref{a}) is equivalent to the curve:
$
y^2  = x^3  - \frac{c_4}{48} x - \frac{c_6}{864}
$
with 
$c_4 =16(a_2^2 - 3 a_4)$ and
$c_6 = -64 a_2^3 + 288\ a_2\ a_4 - 864\ a_6$.}
\(
y^2 = x^3 + a_2 x^2 + a_4 x + a_6,  
\label{a}
\)
where $a_2, a_4,a_6$ are polynomials in $u$ and in the
masses  $m_i$. They are also polynomials in the
scale  $\Lambda$ of the theory for $N_f < 4$,  or
of certain modular functions $e_i(\tau_0)$ for $N_f=4$ or
for ${\cal N}=4$, where $\tau_0$ 
is the coupling as measured at $u=\infty$ in the $N_f=4$ or ${\cal N}=4$ theory.

\section{The elliptic curve in F-theory}
F-theory emerged \cite{Vafa} in order to explain geometrically the
axion-dilaton combination in type IIB string theory in the Kaluza-Klein
spirit as moduli of an internal torus. These moduli are not fixed but 
vary. One takes this to be an elliptic fibration over a base manifold. The
complex structure modulus $\tau=\chi +ie^{-\phi}$ of the fiber then varies
on the upper half plane. Consequently, as we move along closed cycles of
the base, $\tau$ will undergo nontrivial $SL(2;\ZZ)$ transformations.

\vspace{3mm}
Let us see how F-theory is related to M-theory.
Recall that we have that the interpretation of type IIB in terms of F-theory
requires the elliptic curve to have a modulus that varies on a base, which is
taken to be ${\CP}^1=S^2$ for two-folds. 
To relate these F-theory compactifications to other
string theories, one can further compactify on $S^1$ with radius $r_{IIB}$.
This is equivalent to M-theory compactified on a two-torus $T^2$ \cite{Asp}.
Taking the tenth and eleventh radii to be $r_9$ and $r_{10}$ respectively,
the relation is
\begin{equation}
r_{IIB}=(r_9r_{10})^{-3/4}.
\end{equation}
So small type IIB $S^1$ corresponds to large area $T^2$ in M-theory, and 
so the supergravity approximation can be trusted.

\vspace{3mm}
One can also consider what happens to the modular parameter $\tau$.
Since the coupling in type IIB is given by 
\begin{equation}
g_{IIB}=(r_{10}/r_9)^{1/2},
\end{equation}
$\tau$ will capture the conformal class of the metric
on $T^2$, which is equivalent to specifying a holomorphic structure 
\cite{Morrison}. So there is a duality in seven dimensions between 
F-theory on $K3\times S^1$ and M-theory compactified on a four-manifold 
which is fibered by $T^2$'s with
holomorphic structure dictated by $\tau(z)$.
The four-manifolds of this kind are not unique, but there
{\it is} a unique one for which the holomorphic fibration
has a holomorphic section \cite{Morrison}.  
The above suggests the duality\cite{Vafa, Sen1} between 
M-theory on elliptically fibered manifold with a holomorphic 
section and F-theory from $\tau(z)$, further compactified on $S^1$.
A Wilson line has to be turned on along the $S^1$ if the four-manifold
has no holomorphic section \cite{Triples, Morrison}.

\section{Relating the curves}

Motivated by Heterotic/F-theory duality, Sen \cite{Sen1} studied F-theory 
over a
$K3$-surface in the special point in the $K3$ moduli space where the 
surface is the $\ZZ_2$-orbifold of the four-torus $T^4$, 
corresponding to the case where the 
axion-dilaton modulus is constant. Such a background was identified with 
an orientifold of type IIB string theory. This configuration 
of F-theory on $K3$ is T-dual to type I string theory 
on $T^2$, which in turn is equivalent to heterotic string theory on $T^2$.
This provided an embedding in F-theory of Seiberg-Witten theory,
i.e. $N=2$ SYM with group $SU(2)$ and with four quark flavors, whose 
moduli space can be characterized by a gauge-invariant quantity $u$, and 
the complex coupling constant $\tau$ that varies as one moves in the 
$u$-plane \cite{SW1,SW2}. Sen \cite{Sen1} has shown that this is identical 
to the F-theory background with $u$ labelling the coordinates of the base 
of the elliptic fibration and $\tau$ denoting the axion-dilaton modulus. 
And so, for  constant $\tau$ in both cases, one has a constant coupling in 
SW theory and a constant axion-dilaton pair in string theory. The gauge theory solution provides the dependence of $\tau$ on $z$ and the 
parameters. Furthermore \cite{Sen1}, the masses of BPS states can be expressed in 
F-theory in terms of period integrals of the holomorphic 2-form on the 
$K3$-surface. The BPS states become massless precisely when one or more 
of these integrals vanishes and so the surface becomes {\it singular}. 

\vspace{3mm}
An elliptically fibered $K3$ surface can be constructed out of the
Weierstrass form of an elliptic curve by letting the parameters $a_i$  
become polynomials in the $S^2={\CP}^1$ coordinate $z$, and thus has the form
\(
y^2=x^3 + f(z)x + g(z)
\)
where $x,y$ and $z$ are coordinates on the base ${\CP}^1$, and $f(z)$ and  
$g(z)$ are polynomials in $z$ of degree 8 and 12 respectively. This 
describes a torus for each point on ${\CP}^1$ labelled by the coordinate 
$z$. The modular parameter $\tau(z)$ of the torus is determined in terms 
of the ratio $f^3/g^2$ through the relation to the $j$-invariant
\(
j(\tau(z))=\frac{4.(24f)^3}{27g^2+4f^3}.
\label{j}
\)
The compactification of F-theory on this particular $K3$-surface 
corresponds to compactification of type IIB string theory on ${\CP}^1$
labelled by $z$, with the modular parameter given by the axion-dilaton 
pair, 
\(
\tau(z)=\chi(z) + ie^{-i\phi(z)/2}.
\)

\vspace{3mm}
Sen also gave the description in terms of D-branes.
From this point of view, such a background corresponds to a 
configuration of 24 D7-branes transverse to the ${\CP}^1$ and situated at 
the zeroes of the discriminant 
$\Delta \equiv 4f^3 + 27g^2$. In terms of the positions $z_i$ of the 
D7-branes, this is $\Delta=\prod_{i=1}^{24}(z-z_i)$. These singular fibers 
correspond to $j(\tau(z_i))\rightarrow \infty$.  
Then considering a special point in the moduli space where $\tau(z)$
is independent of $z$ requires from (\ref{j}) that the ratio $f^3/g^2$ is 
a constant, which means that $f$ and $g$ can be written in terms of one 
ploynomial $\phi$ of degree 4 in $z$ as $g=\phi^3$ and $f=\alpha \phi^2$,
where $\alpha$ is a constant. 
From the point of view of D-branes, this corresponds to grouping the D7-branes into 4 
sets of 6 coincident D7-branes, situated at the points $z_i$ 
$i=1,\cdots,4$ where $\phi$ vanishes. The effect of this is that 
\cite{Sen1} the resulting base is the orbifold $T^2/I_2$ of 
the two-torus $T^2$ by the group that inverts the signs of both 
coordinates. This comes from an $SL(2;\ZZ)$ monodromy ${\rm diag}(-1,-1)$
around each of the points $z_i$. 

\vspace{3mm}
The above transformation can be identified \cite{Sen1} with the discrete transformation
$(-1)^{F_L} \Omega$ of type IIB, where $(-1)^{F_L}$ changes the sign of 
all the Ramond sector states on the left moving sector and $\Omega$ is the 
orientation-reversal transformation that exchanges the left- and the right 
moving modes on the worldsheet. This means that one is considering type 
IIB string theory compactified on $T^2/I_2$ such that when one moves once 
around each point on $T^2/I_2$ the theory comes back to itself transformed 
by the symmetry $(-1)^{F_L}\Omega$. In other words, the theory can be 
identified with type IIB on $T^2$, modded out by the $\ZZ_2$ 
transformation $(-1)^{F_L}\Omega I_2$.
This is a type ${\rm I}^{\prime}$  orientifold \cite{GP,PCJ} which is related to type 
I theory
by a T-duality transformation. By making an $R \rightarrow 
1/R$ transformation on both circles of $T^2$ one maps the 
$\ZZ_2$-transformation $(-1)^{F_L} \Omega I_2$ to the transformation 
$\Omega$ \cite{Wi, Sen1}.  
The corresponding 
D7-brane configurations is identical to the one obtained from F-theory.
This in turn establishes \cite{Sen1} the conjectural \cite{var, Dab, hull, PW} 
equivalence to heterotic string 
theory on $T^2$.

\vspace{3mm}
Note that there are also other branches of constant $\tau$. In addition to 
Sen's case considered above, there are two other branches \cite{DM}.
First, $f=0$, $g \neq 0$, corresponding to $\alpha \rightarrow 
\infty$, and second, $f\neq 0$, $g=0$, corresponding to $\alpha 
\rightarrow 0$. The constant value of $\tau$ is $i$ for the first case 
and $\exp(i\pi/3)$ for the second case. The $j$-invariant is $13824$ 
for the first case and $0$ in the second case.
These corresponding discriminants are $\Delta=4f^3$ and 
$\Delta=27g^2$, respectively. The periods of $K3$ will then be given by
\(
\Omega_i=C(\tau_i)\int \frac{dz}{\Delta^{1/12}},
\)
where $C(\tau_i)$ are constants corresponding to the constant values
$\tau_i$ equal to $\tau_0$, $i$ and $e^{\frac{i\pi}{3}}$.
Going back to the orbifold picture, these correspond to the $K3$ 
being $T^4/\ZZ_n$ where $n=4$ in the first branch and $n=3,6$ in the 
second branch \cite{DM}. 
It is interesting that singularities of 
type $E_6$, $E_7$ and $E_8$ were obtained for $n=3,4$ and $6$ respectively,
which suggests that a description through weakly coupled D-branes is not possible even in a
limit \cite{DM}.

\vspace{3mm}
We will look at the picture in terms of the moduli space of $K3$.
The $\Z_n$ orbifold construction of $K3$ can be described as follows \cite{Wa}.
Consider a four-torus $T$, where
$T=T^2\times \widetilde{T}^2$  with two
$\ZZ_n$ symmetric two-tori $T^2=\CC/L,\,\widetilde{T}^2=\CC/\widetilde{L}$
which need not be orthogonal.
Let $\zeta\in\ZZ_n$  act algebraically on
$(z_1,z_2)\in T^2\times \widetilde{T}^2$ by $(z_1,z_2)
\mapsto(\zeta z_1,\zeta^{-1}z_2)$. Next mod out this symmetry and blow up 
the resulting singularities; that is, replace each singular point by a chain 
of exceptional divisors, which in the case of $\ZZ_n$-fixed points have as intersection
matrix the Cartan matrix of  $A_{n-1}$. In particular,
the exceptional divisors themselves are
rational curves, i.e. holomorphically embedded
spheres with self intersection number $-2$.
For $n\in\{2,3,4,6\}$ this procedure changes the
Hodge diamond by
$$
\begin{array}{ccccc} &&1\\ &2&&2\\ 1&&4&&1\\&2&&2 \\ &&1 \end{array}
\longmapsto
\begin{array}{ccccc} &&1\\ &0&&0\\ 1&&20&&1\\&0&&0 \\ &&1 \end{array}
$$
and indeed produces a $K3$ surface $X$.
One also obtain a rational map from  $T$ to $X$ of degree $n$ by
this procedure. What is interesting is that the values of $n$ in this construction 
are exactly the ones corresponding to the orbifold limit \cite{Sen1,DM}, 
which makes the above picture consistent. Note that corresponding to the 
above geometric orbifolds are orbifold conformal field theories, whose moduli 
spaces are constructed in \cite{NW}.

\section{Discussion and proposals}
{\bf Orbifolding and the $w_4$ condition}

\vspace{1mm}
\noindent We would like to explain the appearance of the $w_4$ condition in the
analysis of \cite{KS1} in a way that makes it seem less foreign. There, one
condition for constructing the mod 2 part of the elliptically refined
partition function of type IIA string theory was the vanishing of the   
fourth Stiefel-Whitney class of spacetime $w_4=0$. It is known that this
condition is related to type I and heterotic theories rather than
type II string theories. So, if we interpret the elliptic curve of
elliptic cohomolgy as that of F-theory, and we further go to the orbifold
limit of Sen,
then we naturally connect to the latter two string theories.
This more explicitly explains the proposal in \cite{KS1} that we see a
unification of the various string theories when viewed through the eye of
elliptic cohomology.   

\vspace{3mm}
{\bf  Explicitly relating the curves}\\
There are three elliptic curves considered in this note. Our arguments in this paper were 
mostly on the relation between the two `physical' elliptic curves (defined over $\CC$), namely the Seiberg-Witten 
and the F-theory curves. This is just Sen's picture of the orbifold limit of $K3$. 
Combined with the previous work \cite{KS1,KS3} this suggests that all three curves are related
in a precise way. However, we did not make any quantitative claims or checks in this note. 
One way of doing so is to relate the corresponding coefficients of the three elliptic curves. 
In Seiberg-Witten theory, the $a_i$ are given by polynomials in the masses of the BPS states, 
and hence in charges for those states. This is related to BPS states in the M-brane configurations. It would be very interesting to make such a match quantitatively, and to (re)produce formulae for 
BS states.  Sen has shown that there is a consistent map between
the orientifold and the F-theory,  $m_i=c_i$,$1\le i\le 4\ $, where 
$m_i$ label the F-theory and the parameters $c_i$ label the orientifold.

\vspace{3mm}
{\bf The $q$-expansions}

\vspace{1mm}
\noindent The interpretations in this note also unify the emerging loop expansions in 
the M2-M5 system ( and hence 
in the gauge theory) and in the F-theory picture.
What we are advocating is that the generalized 
cohomology theories that we are dealing with have the correct mathematical 
structure to incorporate the interactions between M2-branes and M5-branes,
which, in light of the arguments in \cite{DVV}, means that this also incorporates
bound states and interaction of type IIA string theory (and type IIB 
by T-duality). In particular, this suggests identifying the parameter $q$ that appeared in 
the refinement of the type IIA and M-theory partition functions in \cite{KS1}
as a coupling constant which encodes the interactions of M2-M5 systems. Then, this 
would also imply that this coupling constant also encodes the interactions
of strings and branes in ten dimensions. This suggests the picture
\(
\begin{CD}
K(X) @> {\rm strong~coupling}>> 
K[[q]](X)
\end{CD}
\)
in such a way that taking the weak coupling limit just sends $q
\rightarrow 0$,
and we go back from the RHS to the LHS. The interpretation of $q$ as a
coupling constant is perhaps close to that in topological string theory. In particular,
it would be interesting to provide a connection to Gromov-Witten invariants.

\vspace{3mm}
{\bf Reduction of coefficients}

\vspace{1mm}
\noindent One might use the standard procedure in (mathematical) gauge theory of working 
over finite fields and then translating back to the 
base field considered. \footnote{One might argue that, in our context, this is more than just a tool
since the finite fields appear in some versions of elliptic cohomology.}
Starting from the elliptic curve $\cE$, 
we get the reduced curve  $\overline{\cE}$ where the coefficients $a_i$ are reduced to 
${\overline{a}}_i$ that take values in the finite field, via what is called the reduction homomorphism
$ r_p$ from the original field to the finite field. The discriminant of ${\overline{\cE}}$
is ${\overline{\Delta}}$, the reduction mod $p$ of the discriminant
$\Delta$ of $\cE$. Clearly ${\overline{E}}$ is nonsingular if and only if
 ${\overline{\Delta}}\neq 0$.
 The elliptic curves are divided into curves with bad reduction and curves 
 with good reduction, depending on whether the discriminant of the reduced 
 curve is  zero or not.
  $\cE$ has {good (bad) reduction at $p$ provided that
$\overline{\cE}$ is nonsingular (singular) at $p$. In the former case, the
reduction function $r_p$
is a group morphism.  In the latter this means that $\overline{\cE}$ is not an
elliptic curve. Further, $\cE$ has additive reduction provided 
that the singularity in $\overline{\cE}$ is a cusp, and $\cE$ has
multiplicative reduction provided that $\overline{\cE}$ has a node.
What is the point of working over finite fields? This is part of some deep connections to
number theory.
However, for us we are just using the correspondence between theories
defined on continuous base and theories defined over finite fields. This is quite useful in 
other situations, for example in gauge theory, where striking analogies are drawn between the two pictures, as well as the utility of working over the finite fields to perform 
explicit calculations of physically interesting quantities.

\vspace{3mm}
{\bf Deforming away from orbifold limit}\\
In this paper we have restricted to constant value of the modular parameter. This fits nicely with the elliptic cohomology picture. However, the physics allows for more, namely for $\tau$ which is a field, which can be viewed as a deformation away from the orbifold limit \cite{Sen1}. In F-theory such deformations correspond to splitting the 6 coincident 
zeros of $\Delta$ away from each other. From the orientifold point of 
view, this corresponds to moving the 4 coincident D7's away from the 
orientifold plane. Curiously, in the presence of D7-branes, $\Delta$ is not inverted because the positions of the D7-branes correspond to zeros of $\Delta$; so one 
has to take that into account in the elliptic cohomology picture, i.e. if $\Delta$ is inverted in 
$E$ then the theory cannot contain D7-branes, and conversely, if we want to describe 
D7-branes together with the singularities then we cannot invert $\Delta$. 
The asymptotic value of the variable $\tau$ is taken to 
be the special value $\tau_0$. The deformation away from the orbifold 
limit is described by a surface of the form
\(
y^2=x^3+{\widetilde f}(z) x + {\widetilde g}(z),
\)
where now ${\widetilde{f}}$ and ${\widetilde{g}}$ are polynomials in $z$ 
of 
degree 2 and 3 respectively. This gives five complex paramters after 
removing one by an overall shift of $z$ and another by a rescaling of $x$ 
and $y$. The corresponding five parameters in Seiberg-Witten theory are 
given by the 4 quark masses $m_i$ ($i=1,\cdots,4$) and the complex 
coupling constant $\tau_0\equiv (\frac{\theta}{2\pi} + \frac{4\pi 
i}{g^2})$. The value of $\tau$ is \cite{Sen1}
\(
 \tau(z)=\tau_0 + \frac{1}{2\pi i} \left(\sum_{i=1}^{4} \ln(z-z_i)-4\ln(z) 
\right),
\label{tau}
\)
where $z_i$ is $m_i^2$ or $c_i^2$ for Seiberg-Witten and the orientifold 
theory respectively

\vspace{3mm}
In this respect then, the full physical theory requires more than elliptic cohomology,
i.e. something that has more aspects of $K3$. 
We might say that elliptic cohomology is at a special point in the moduli space 
of $K3$, and the required theory contains a generalization of the elliptic curve.
Such a generalization would be a cohomology theory built out of $K3$, e.g. a 
$K3$-cohomology.
From considerations in arithmetic 
algebraic geometry, one can build formal group laws corresponding to $K3$ , and so in principle 
such a theory could exist. However, we do not know of a construction. 
In fact Sen's arguements can be generalized from $K3$ to higher dimensional 
Calabi-Yau spaces \cite{Sen2, Sen3}, and thus hints at room even for Calabi-Yau 
cohomology (beyond $K3)$.  Further, at 
weak coupling and even away from the special point in the moduli space, the 
orientifold and the F-theory descriptions coincide. However, at strong 
coupling the orientifold description breaks down near the orientifold 
point. This implies that \cite{Sen1} F-theory provides the correct 
description of the background field configuration of this theory, and the 
orientifold background must be modified by {\it quantum corrections} so as 
to coincide with the F-theory background, and the description is 
nonperturbative since one has effects of the form $\exp(i\pi \tau_0/2)$.
Consequently, the full F-theory limit was proposed to describe the 
quantum-corrected version of the orientifold background. 
For us, this suggests the conjecture that in order to capture the full quantum theory, 
elliptic cohomology needs to be corrected by some higher cohomology theory, which is either
a version of elliptic cohomology adapted to elliptic fibrations i.e. which encodes variation in the moduli space, or a form of Calabi-Yau cohomology.

\vspace{5mm}
{\large \bf Acknowledgements}

\vspace{1mm}
The author thanks Wolfgang Lerche for very useful discussions, encouragement and hospitality at CERN,  Matilde Marcolli for useful discussions and hospitality at MPI-Bonn, and the organizers 
of the program ``Mathematical Structures in String Theory"  and the KITP for their hospitality.
He also thanks Mike Douglas and Tony Pantev for helpful discussions, Greg Moore and Emil Martinec
for useful questions and comments, and Michael Atiyah and Raoul Bott for insightful remarks. Thanks are due to Arthur Greenspoon for useful comments and suggestions on improving the presentation.
This research was supported in part by the National Science Foundation under Grant No. PHY99-07949.  


\end{document}